# Beyond the upper limit of magnetic anisotropy in two-dimensional transition metal dichalcogenides

Dorj Odkhuu

*Department of Physics, Incheon National University, Incheon 406-772, Republic of Korea*

Exploring an upper limit of magnetic anisotropy in two-dimensional materials, such as graphene and transition metal dichalcogenides, is at the heart of spintronics research. Herein, an atomic-scale perpendicular magnetic anisotropy up to an order of 100 meV per atom, which is far beyond the ordinarily obtained value in graphene and pristine transition metal dichalcogenides, is demonstrated in individual ruthenium and osmium adatoms at a monosulfur vacancy in molybdenum disulfide. We further propose a phenomenological model where a spin state transition that involves hybridization between molybdenum $a_1$ and adatomic $e'$ orbitals is a possible mechanism for magnetization reorientation.

Single-molecule magnets have great potential for quantum spin processing or spin-dependent electronics because of their small size and intrinsic magnetism. In particular, individual magnetic atoms can meet today's demand to maximize the information storage density in a bit unit area, which requires breakthroughs that lead to the atomic-scale upper limits of magnetic anisotropy [1]. However, the concept of magnetocrystalline anisotropy, which is a preferential spatial orientation of magnetization, is not relevant in an individual isolated atom; i.e., magnetic moments freely rotate in any direction without energy cost. Thus, the upper limit of magnetic anisotropy of a single atom must be realized by properly selecting the contact medium, where the perpendicular magnetocrystalline anisotropy (PMA) is essential [2]. With recent advances in fabrication techniques, one of the pioneering experiments has demonstrated that individual Co adatoms on a Pt (111) surface exhibit a large PMA of ~9 meV per Co atom [3]. Thereafter, there have been a few more subsequent experimental reports on an individual magnet's anisotropy; for example, unexpectedly large anisotropies up to the upper possible limit of a 3$d$ metal atom, ~60 meV per atom, have been achieved in Fe and Mn adatoms on a CuN surface [4] and Co on MgO (001) [5]. Although these remarkable findings provide a new concept to explore atomic-scale anisotropy, the renewal of research targets seemingly resides in the use of even two-dimensional materials such as graphene [6,7] and possibly transition metal dichalcogenides.

Herein, we report results of first-principle calculations on an atomic-scale PMA up to an order of 100 meV per Os atom, which was identified in individual transition metal (TM = Ru and Os) atoms placed at the sulfur vacancies in molybdenum disulfide ($MoS_2$), far beyond the ordinarily obtained values in graphene and pristine transition metal dichalcogenides. Furthermore, as a phenomenological model, we propose that such a large PMA undergoes a transition into an easy magnetization axis in plane at the low-spin (LS) → high-spin (HS) state transition. The ligand field and single-particle energy spectra analyses with spin-orbit Hamiltonian matrix elements reveal that such strong hybridization creates large spin-orbit coupling pairs between the adatomic nondegenerate $e'$ states $d_{xy}$ and $d_{x^2-y^2}$, which determines the magnetic anisotropy of transition-metal-doped molybdenum disulfide.

Density functional theory (DFT) calculations were performed using the Vienna *ab-initio* simulation package (VASP) [8] with the generalized gradient approximation (GGA) for the exchange correlation functional [9]. The modeled geometry shown in Fig. 1(a) contains a single TM magnet adsorbed on a 3x3 lateral unit cell of $MoS_2$ and a vacuum region of 15 Å along the z-axis. The most stable adsorption site of Ru and Os adatoms on non-defective $Mo_2$ is the Mo-top site [Fig. 1(b)]. In Fig. 1(c), TM adatoms are embedded into an intrinsic defect of a monosulfur vacancy (defective), which occurs notably often during the sample growth [10]. The latter geometry provides the LS state complex in a trigonal bipyramidal-like crystal field for TM adatoms. The energy cutoff of 400 eV and a 9x9x1 $k$-mesh were imposed for the lattice and ionic relaxations, where the forces acting on each atom were less than $10^{-2}$ eV/Å. The spin-orbit coupling (SOC) term was included using the second-variation method employing the scalar-relativistic eigenfunctions of the valence states [11]. The magnetic anisotropy energy (MAE) is obtained based on the total energy difference when the magnetization directions are in the xy-plane ($E^{\parallel}$) and along the z-axis ($E^{\perp}$), MAE = $E^{\parallel} - E^{\perp}$, where convergence with respect to the $k$-point sampling is ensured.

In Table I, the optimized in-plane lattice (3.18 Å) of pristine $MoS_2$ is consistent with the reported theoretical [12] and experimental results [13] and almost remains unchanged upon the adsorption of TM atoms.

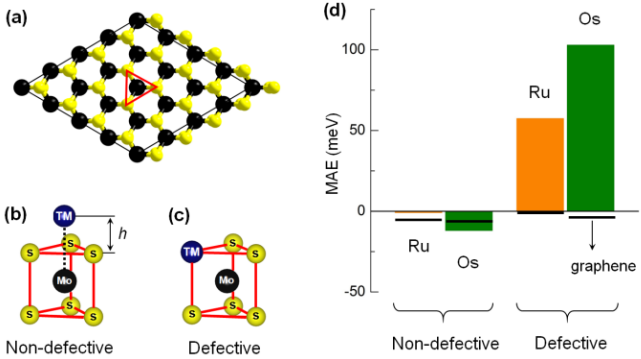

FIG. 1: (a) Top view of a single-layer of MoS$_2$. The large black and small orange spheres represent Mo and S atoms, respectively. The most stable binding sites of the transition metal (TM) adatom (large red sphere) on MoS$_2$ (b) without (non-defective) and (c) with (defective) a sulfur vacancy. (d) MAE of individual Ru (orange) and Os (green) adatoms on non-defective (left) and defective (right) MoS$_2$. The corresponding results for graphene are also illustrated in lines for comparison.

In general, the calculated binding energies, which are formulated as $E_b = E(h_{equilibrium}) - E(h_{isolated})$, where $h$ is the height of an adatom with respect to the top sulfur plane, are negative (Table I). This value implies the favorable formation of TM-Mo$_2$ pairs. In particular, the binding affinity is more favorable when the adatoms sit at the sulfur-site vacancy in defective MoS$_2$, as found in Fe and Mn [14]. The magnetic energy gain $\Delta E > 0$, which is the energy difference $\Delta E = E_{NSP} - E_{SP}$ between the spin-polarized (SP) and non-polarized (NSP) phases, indicates the feasibility of induced magnetism in TM adatoms, as shown in Table I. This magnetic instability occurs because of the band narrowing in the reduced dimension, which enhances the densities of state (DOS) at the Fermi level, thereby satisfying the Stoner criterion of unity. Notably, both Ru and Os adatoms exhibit larger spin moments and $\Delta E$ on defective MoS$_2$ than on non-defective MoS$_2$ mainly because of the strong hybridization between Ru 4$d$ (Os 5$d$) and Mo 4$d$ orbitals.

Figure 1(d) shows the calculated MAE of Ru and Os adatoms on non-defective and defective MoS$_2$. For both TMs, the non-defective and defective systems exhibit distinct trends in MAE: negative and positive in sign, respectively. The latter stands for the preferable direction of magnetization normal to the MoS$_2$ plane, i.e., PMA. Noteworthy, PMA increases severely from 3$d$ Fe (~9 meV) to Ru (~57 meV) and eventually reaches an enormously large MAE of ~102 meV per Os atom.

TABLE 1: Optimized in-plane lattice $a$ and adatom height with respect to the top sulfur plane $h$ (Å), binding energy $E_b$ and magnetic energy $\Delta E$ (eV), spin moment $m_s$ and orbital moment difference $\Delta m_o$ ($\mu_B$), and charge transfer $\Delta \rho$ ($e$) of Ru and Os adatoms on non-defective and defective MoS$_2$.

| TM | $a$ | $h$ | $E_b$ | $\Delta E$ | $m_s$ | $\Delta m_o$ | $\Delta \rho$ |
|---|---|---|---|---|---|---|---|
| Non-defective MoS$_2$ | | | | | | | |
| Ru | 3.19 | 1.09 | −3.62 | 0.28 | 1.08 | 0.06 | −0.57 |
| Os | 3.19 | 0.98 | −2.94 | 0.22 | 0.99 | 0.20 | −0.64 |
| Defective MoS$_2$ | | | | | | | |
| Ru | 3.17 | 0.28 | −5.21 | 0.47 | 1.69 | −1.07 | 0.08 |
| Os | 3.18 | 0.20 | −4.84 | 0.34 | 1.59 | −0.63 | 0.41 |

The spin-orbit effect as a physics origin of anisotropic phenomena is simply justified because of the inherently large SOC of 4$d$ and 5$d$ orbitals [15-17]. However, when the $d$–$d$ bands become more overlapped with metallic bonding, the strengthened crystal field effect plays a more important role to realize such a large PMA. To support this scenario, we perform similar calculations on non-defective and defective graphene, where one carbon atom in a 4x4 unit cell of graphene is replaced by a TM adatom. Our results show that Ru and OS adatoms have MAE = −5.9 (−0.5) and −6.1 (−3.9) meV on non-defective (defective) graphene, as illustrated in Fig. 1(d), and all of them prefer an easy magnetization axis in plane, which is comparable to non-defective MoS$_2$ with the $p$–$d$ covalency.

From a practical viewpoint, the sufficiently large PMA at the atom length scale of TM-doped MoS$_2$ is worth noting to prevent a stable magnetization axis from thermal fluctuations [2]. The thermal stability factor $\Delta$ is maintained by the large PMA through $\Delta = KV/k_BT$, where $K$, $V$, $k_B$, and $T$ are the anisotropy, volume, Boltzmann constant, and temperature, respectively [18]. Moreover, this drastic sensitivity of magnetic anisotropy in terms of the sign suggests another viewpoint: an engineering of magnetization orientation through a surface treatment. For example, an in-plane magnetization is favored if one considers one- or two-side neighbor bisulfur vacancies (results not shown).

We further inspect the relationship between the orbital moment $m_o$ and MAE according to Bruno's model [19]: MAE = $-\xi/4\mu_B \Delta m_o$, where $\xi$ is the strength of SOC, and $\Delta m_o = m_o^{\parallel} - m_o^{\perp}$. The calculated $\Delta m_o$ of Ru and Os adatoms are shown in Table I, where $\Delta m_o > 0$ ($\Delta m_o < 0$) on non-defective (defective) MoS$_2$. These results adequately obey the Bruno relation: the easy magnetization axis coincides with the direction that has the largest orbital moment.

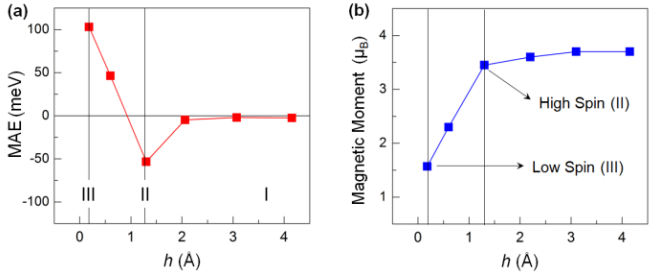

FIG. 2: (a) MAE and (b) spin magnetic moment of the Os adatom on defective $MoS_2$ as a function of the Os-$MoS_2$ separation $h$. I, II, and III denote the HS states of the Os adatom at the isolated ($h > 3$ Å) and intermediate ($h = 1.3$ Å) phases and the LS state at the stable phase on $MoS_2$ ($h = 0.2$ Å), respectively.

Naively, one can note that the Bruno model is illustrative in the analysis of an individual magnet's anisotropy, which is associated with the large ξ and $\Delta m_o$.

Next, to unveil a determinant role of the crystal field on magnetic anisotropy, we plot the MAE and spin magnetic moment of the Os adatom on defective $MoS_2$ as a function of $h$ in Figs. 2(a) and 2(b), respectively. In a free-atom region (phase I), $h > 3$ Å, and MAE is negligible, which is trivial and implies an easy spin axis in random directions unless the Zeeman effect occurs. When the adatom moves toward $MoS_2$, phase I → phase II, and MAE becomes negative (i.e., easy magnetization axis in plane) and reaches a maximum of ~ −50 meV at $h \approx 1.3$ Å. Note that MAE at $h \approx 1$ Å, which corresponds to the equilibrium binding distance of the Os adatom on non-defective $MoS_2$, almost reproduces the value of −12 meV found for the non-defective case [Fig. 1(d)]. Furthermore, MAE changes in sign to positive when $h$ is varied from 1 to 0.2 Å (HS → LS or phase II → phase III), and the PMA is restored with MAE = ~102 meV at the stable $h = 0.2$ Å of the LS state complex.

Meanwhile, the spin moment of free Os at phase I, when $h > 3$ Å, should be governed by the unpaired electron spin counts according to Hund's rule, which is 4 $\mu_B$ for the $d^6$ state. As shown in Fig. 2(b), the calculations yield a slightly smaller magnetic moment of approximately 3.7 $\mu_B$, which is acceptable in at least the standard DFT limits. This value almost remains unchanged (but slightly decreases) at $h \approx 1.3$ Å (phase II), where MAE ≈ −50 meV, which indicates the HS state with five majority electrons in nondegenerate $e''$, $e'$, and $a_1$ states and one minority electron in the lowest-level $e''$ state, which is split by the weak crystal field. The spin crossover and moment can be estimated using the average spin state formalism, $\bar{S} = S_{LS}G_{LS} + S_{HS}G_{HS} e^{-\Delta E/k_B T} / G_{LS} + G_{HS} e^{-\Delta E/k_B T}$, where $G_{LS}$ and $G_{HS}$ are the degeneracies of the LS and HS states, respectively. If these degeneracies occur only because of the spin, $G = 2S + 1$. $\Delta E$ is the energy difference between the LS and HS states, which is notably small for the present case. From $\bar{S} = 5/3$ with $S_{HS} = 2$ and $S_{LS} = 0$, the spin moment at the spin crossover is 3.33 $\mu_B$, which is almost reproduced in our calculation (3.38 $\mu_B$) at $h \approx 1.3$ Å. The magnetism onset in reverse occurs immediately at this intermediate phase with $S = 5/3$ and decreases sharply when $h$ is further reduced. The calculated spin moment at $h = 0.2$ Å or phase III is 1.59 $\mu_B$ [1.69 $\mu_B$ for Ru (Table I)], which is the feature of the LS state complex with $S = 1/2$, not $S = 0$. The LS instability is caused by the stronger crystal field effects as a result of the $d$-shell overlapping with surrounding Mo-$d$ orbitals. This result is indeed consistent with the effective magnetic moment of 1.73 $\mu_B$ in the LS state complex calculated from the spin-only formula $\mu_{eff} = \sqrt{n(n+2)}\ \mu_{eff}$, where $n$ is the number of unpaired electrons.

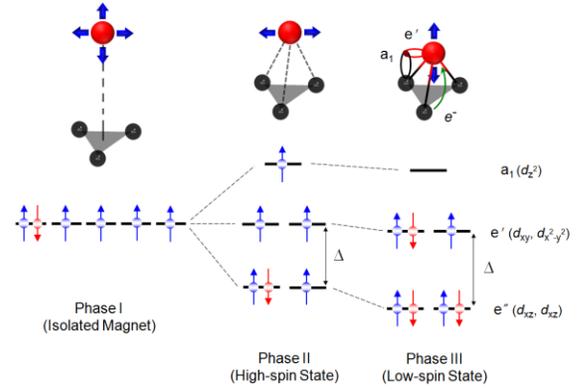

FIG. 3: Schematic diagram of the ligand field splitting of the high-spin and low-spin complexes in a trigonal coordinate. The red and black spheres represent the Os adatom and Mo atoms in MoS2, respectively. When the adatom approaches $MoS_2$, five d orbital states split into two doublets $e'$ ($d_{xy}$ and $d_{x^2-y^2}$) and $e''$ ($d_{xz}$ and $d_{yz}$), and one singlet $a_1$ ($d_{z^2}$). Their electron occupancies and relative energy levels are shown depending on whether high- or low-spin state. The thicker blue arrows denote the spin direction of the Os adatom. In the low-spin state complex, hybridization between the Mo $a_1$ and Os $e'$ orbitals and electron transfer from the Mo atoms that occupies the Os $e'$ orbital level are also indicated.

Small spin moments are beneficial in the context of spintronic applications, which is the key factor to reducing the current density for magnetization switching [17]. The switching current $I_c$ is expressed as $I_c = \frac{2e}{\hbar}\frac{\alpha}{\eta}M_S V(H_k + 2\pi M_S)$, where $\alpha$, $M_S$, and $H_k$ represent the Gilbert damping coefficient, saturation magnetization, and Stoner-Wolfarth switching field, and $\eta$ is the spin polarization factor [20]. Because $H_k \ll M_S$ in the present system, $I_c \approx M_S^2 V$. A small $V$ of individual magnets on two-dimensional MoS$_2$ is also favored to lower $I_c$, which is detrimental to the thermal stability $\Delta$, as previously mentioned.

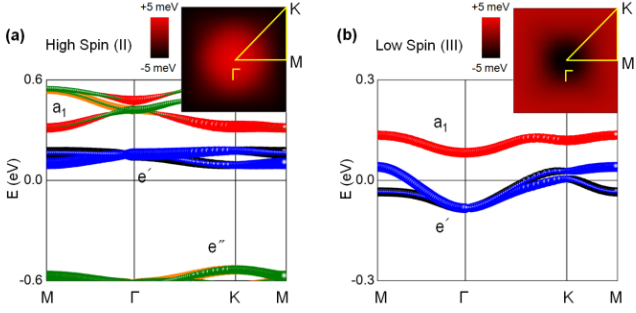

FIG. 4: The minority spin band structure of an Os adatom at (a) $h = 1.3$ Å and (b) $h = 0.2$ Å. The Fermi level is set to zero. The inset shows the MAE distribution over the $k$-space, MAE($k$). The black and red areas represent negative and positive MAE($k$), respectively, with the energy scale of −5 meV to +5 meV.

Because the main physics for the spin state transition mediates the magnetization reversal, a simple schematic level diagram for the Os $5d^6$ orbital splitting in the presence of a trigonal ligand field (by defective MoS$_2$) is sketched in Fig. 3. The spin state transition of a free magnet's degenerate five $d$ orbitals (phase I) into the ground state nondegenerate LS state complex (phase III) can occur in a trigonal bipyramidal-like crystal field after the Os $5d$ and Mo $4d$ hybridization (i.e., strong crystal field) forms, which passes through the intermediate HS state or phase II (i.e., the weak crystal field). Relevant orbitals involved in this crystal field splitting are the low-lying doublet $e''$, high-lying doublet $e'$ and singlet $a_1$ in either phase II or III. However, the relative occupancies of spin-up and -down states, their energy level, and crystal field splitting ($\Delta$) are not identical between phases II and III, which offers more freedom to provide different energy landscapes in the framework of an SOC Hamiltonian matrix, as explicitly addressed in the following discussion.

In the absence of MoS$_2$, the LS state of Os $5d^6$ results in zero net moment because of the complete two electron pairs at the low-lying $e''$ and one at the high-lying $e'$ energy level, i.e., $S = 0$. However, in the presence of MoS$_2$, the level splitting in the LS state complex also involves an interplay between orbital hybridization and charge quantization. From Bader charge analyses, the charge increment of the Os adatom at $h = 0.2$ Å (phase III) with respect to the nominal charge (8 $e$) of a free atom (phase I) is 0.41 $e$, which is involved in the occupation of a high-lying $e'$ level with the majority spin state because of the Pauli exclusion principles, as illustrated in Fig. 3; thus, $S = 1/2$. This charge accumulation is from Mo atoms in MoS$_2$ because Os atoms have larger electronegativity (2.2) than Mo (2.16). This result is also the case for Ru adatoms, whose electronegativity is the same as Os. By contrast, the electrons deplete from TM adatoms to sulfur atoms in non-defective MoS$_2$ because of the larger electronegativity (2.58) of sulfur, as shown in Table I.

The HS → LS state transition evolves in different energy landscapes around the Fermi level, which consequently modulates MAE. To address this physics problem, we follow the recipe of the second-order perturbation theory by Wang et al., where MAE is determined by the SOC between occupied and unoccupied bands [21]: $MAE = \xi^2 \sum_{o,u} \frac{|\langle o|l_z|u\rangle|^2 - |\langle o|l_x|u\rangle|^2}{\varepsilon_u - \varepsilon_o}$, where $o$ ($u$) and $\varepsilon_u$ ($\varepsilon_o$) represent the eigenstates and eigenvalues of occupied (unoccupied) states, respectively. The two spin-channel decomposition terms of MAE, which involve the spin-up state, ↑↑ and ↑↓, are neglected, which is analogous to the series of $d$ transition metal systems explored in the previous full-potential calculations. The positive and negative contributions to MAE are characterized using $l_z$ and $l_x$ operators, respectively.

In Figs. 4(a) and 4(b), we plot the $k$-resolved MAE (MAE($k$)) and the band structures, which were projected onto the Os $d$ orbitals for the HS ($h = 1.3$ Å or phase II) and LS ($h = 0.2$ Å or phase III) complex, respectively. In the insets, MAE($k$) along the high symmetry line (MGKM) of the *hexagonal* Brillouin zone (BZ) qualitatively represents the total MAE for the HS (negative) and LS (positive) states. In particular, the dominant contributions at the M and K points are prominent. In the HS state, the SOC pairs between the filled $e'$ ($d_{xy}$ and $d_{x^2-y^2}$) and empty $e''$ ($d_{xz}$ and $d_{yz}$) bands at M and K provide the negative MAE($k$) therein through $< xy,x^2-y^2 | l_x | xz,yz >$; thus, the total MAE < 0. This negative MAE($k$) decreases around G owing to the enlarged $e''$- $e'$ energy split from $\varepsilon_u - \varepsilon_o = 0.59$ eV at K to 0.78 eV at G, which reduces the second (negative) term of the aforementioned equation. For the LS complex [Fig. 4(b)], it is obvious that the positive MAE($k$) at M and K come from the non-degenerate doublet $e'$ bands across the Fermi level. This positive MAE($k$) from $< xy,x^2-y^2 | l_x | xz,yz >$, which has the largest contribution to the PMA, by a factor of 2 compared to the other matrix elements, disappears at G because of the absence of filled $e''$ bands. Instead, the presence of the singlet $a_1$ ($d_z^2$) state in the empty band leads to the negative MAE($k$) at G. Summing up all these features, the opposite sign in MAE between the HS and LS complexes is due to an interplay of SOC states between $d_{xy}$ and $d_{x^2-y^2}$ across the Fermi level, which is a result of the strong hybridization with the Mo $d_{z^2}$ state.

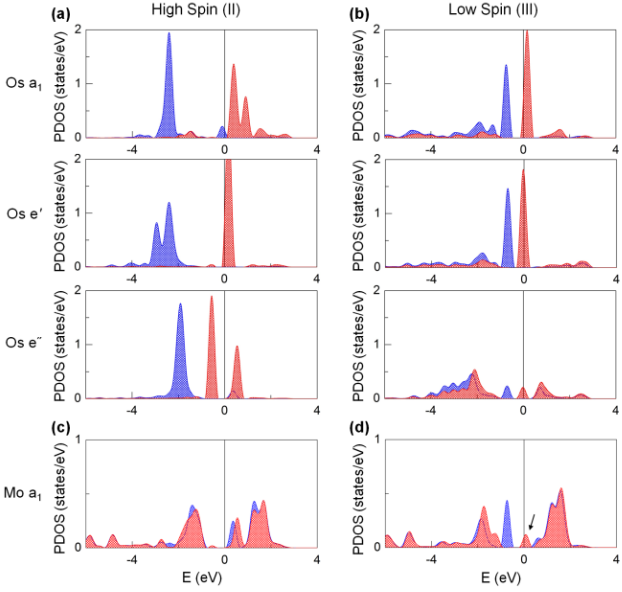

FIG. 5: Spin- and orbital-decomposed PDOS of an Os adatom on defective MoS$_2$ at (a) $h = 1.3$ Å and (b) $h = 0.2$ Å. (c) and (d) The corresponding (total) TDOS of a Mo atom underneath the Os adatom. In (d), the minority spin $a_1$ state is emphasized. The black and red areas denote the spin up and down states, respectively. The Fermi level is set to zero.

To better justify the hybridization-mediated spin state transition, we show the partial DOS (PDOS) of the Os adatom on defective MoS$_2$ in Figs. 5(a) and 5(b) for the HS and LS complexes, respectively. The corresponding total DOSs (TDOSs) of the Mo atom below Os are also plotted in Figs. 5(c) and 5(d). Apparently, five $d$ orbitals in the HS state are fully occupied by the majority spin electrons. While the minority doublet $e'$ and singlet $a_1$ states are empty, the half-filled $e''$ doublet reflects the occupied $d_{xz}$ (or $d_{yz}$) electron in the minority spin channel. When the HS transforms to the LS complex [Fig. 5(b)], a substantial band broadening appears in the $e''$ spectra, indicating that the Os 5$d$ electrons interact with those of MoS$_2$, which are involved with the $d_{xz}$ and $d_{yz}$ orbitals. Notably, the minority spin $e'$ states reside more in the filled bands, which exhibit the feature of common peak structures (hybridization) with Mo $d_{z^2}$ at the Fermi level. The majority spin doublets $e'$ and $e''$ remain filled.

In summary, individual Ru and Os adatoms at the sulfur vacancies in MoS$_2$ have been shown to exhibit a tremendously large PMA up to an order of 100 meV per Os atom. The present study provides a viable route to achieve atomic-scale PMA by carefully exploring the large SOC energy, orbital magnetism, and ligand field for a suitable choice of an individual adatom and underneath two-dimensional substrate pairing. This large anisotropy with low magnetization and small volume is the key factor to realizing materials with a low switching current and high thermal stability for spintronic applications. Moreover, as a phenomenological model, the HS → LS transition, which is associated with the hybridization between Mo 4$d$ and Os 5$d$ (Ru 4$d$) orbitals, is proposed to correlate with the magnetization reversal. Similar results can be expected in other two-dimensional transition metal dichalcogenides.